\documentclass[a4paper]{jpconf}
\usepackage{graphicx}
\begin{document}

\title{CMS RPC data taking during the LHC Run-2 and activities during Long Shutdown 2}

\author{Kevin Mota Amarilo, on behalf of CMS Collaboration}

\address{Rio de Janeiro State University, Institute of Physics, R. São Francisco Xavier, 524, Rio de Janeiro - RJ, 20550-013, Brasil}

\ead{kevin.amarilo@cern.ch}

\begin{abstract}
The CMS experiment collected around 150 fb$^{-1}$ of proton-proton collision data at $\sqrt{s}$ = 13 TeV during the Run-2 data taking period of LHC. The CMS RPC system provided redundant information for robust muon triggering, reconstruction and identification. To ensure stable data taking, the CMS RPC collaboration has performed detector operation, calibration and performance studies. After the end of Run-2, it was started the second LHC long shutdown period (LS2), an important opportunity for maintenance and preparation for the next data taking period (Run-3) and the installation of services in preparation for the Phase-II upgrade. The activities included maintenance of power, gas and online systems. In this presentation, the overall performance of the CMS RPC system during the Run-2 period is summarized as well as all the activities done in preparation for future data taking periods.
\end{abstract}

\section{Introduction}

One of the main actors of the CMS (Compact Muon Solenoid) experiment \cite{CMS_Experiment2008} is its Muon System \cite{CMS_MuonTDR1997}, which uses gaseous detector technologies to accomplish muon triggering, identification, transverse momentum and charge measurement. 

Through the end of Run 2 in 2018, the components of the CMS Muon system were: drift tubes (DT) at barrel region ($|\eta| < 1.2$), cathode strip chambers (CSC) at endcap region ($1.2 < |\eta| < 2.4$) and Resistive Plate Chambers (RPC), in barrel and endcap regions ($|\eta| < 1.9$). During LS2 a new type of detector was installed, the gas electron multiplier (GEM), as part of the Muon System Phase-2 Upgrades to complement measurements in the higher $\eta$ region for Phase-II upgrade \cite{CMS_MuonPhaseII}, which will also include new RPC chambers, the so called improved-RPC (iRPC).

The CMS RPC system is composed of double gap phenolic resin (called bakelite) chambers, operating in avalanche mode with gas mixture 95.2\% $C_2H_2F_4$, to enhance the ionization caused by incident particles, 4.5\% $iC_4H_{10}$ as a quencher gas to reduce streamer formation and 0.3\% $SF_6$ to control secondary ionization. The RPCs are designed and calibrated to have a time resolution of around 2 ns and a number of adjacent strips fired in a single muon hit (Cluster size) between 2 and 3.

\section{CMS RPC Run-2 data taking}

Between 2015 to 2018 during Run-2, CMS recorded proton-proton collisions with $\sqrt{s}$ = 13 TeV from LHC, with total integrated luminosity of 150.26 \; fb$^{-1}$, out of which only 0.15 \% was lost due to RPC problems. The total accumulated charge in Run-1 and Run-2 of the RPC system was 2.3 \; mC/cm$^2$ in barrel and 7.5 \; mC/cm$^2$ in endcap. Studies at the CERN Gamma Irradiation Facility ++ (GIF++) have showed no drop in efficiency of RPCs up to an integrated charge of 153 mC/cm$^2$ \cite{Aging}

\subsection{RPC efficiency and cluster size stability}

Hit efficiency and cluster size (CLS) are important parameters for RPC performance. An RPC hit's geometrical coordinates are calculated as the geometrical center of the cluster formed by the fired adjacent strips. The CLS of RPC hits should be kept less than 4 strips to avoid fake muon triggers, therefore, proper calibration is very important to keep good performance.

The calibration is done by analyzing the efficiency and CLS dependence on the effective high voltage (HV), which is the high voltage applied corrected by the environmental pressure and temperature variations. HV working point scans were performed once or twice a year in dedicated collision runs, effective voltage is applied in values between 8600 and 9800 V. The collected data is analyzed and the proper working points are selected. More information on the calibration can be found in \cite{AbbresciaWP2005}.

\begin{figure}[h]
\centering
\includegraphics[width=.6\textwidth]{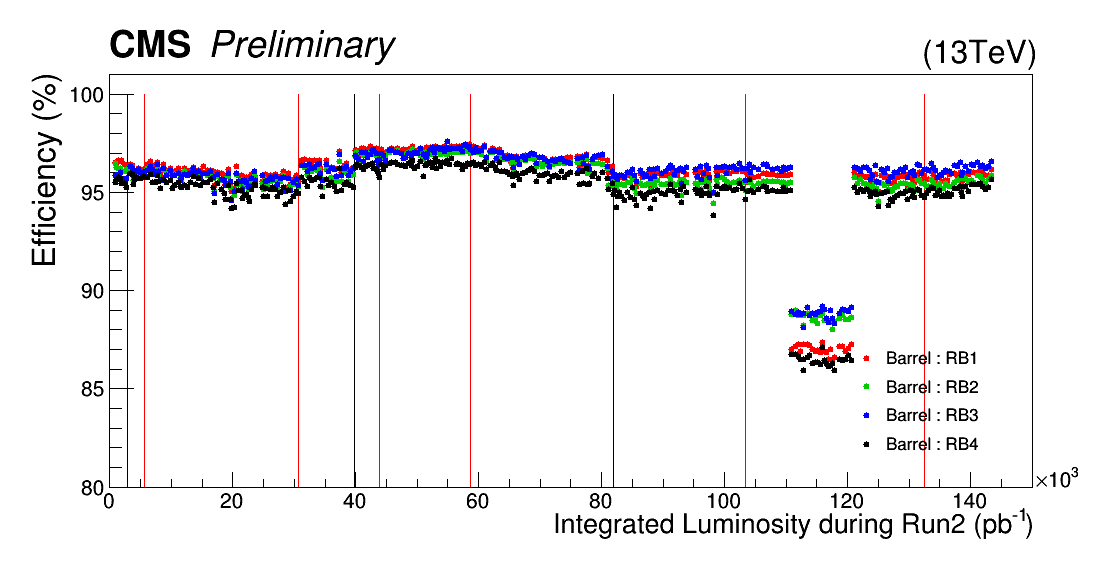}
\caption{\label{BarrelEff}RPC average efficiency vs integrated luminosity during Run-2 for barrel stations. Red vertical lines show the planned technical stops (TS) and the grey ones — Year-End-Technical stops (YETS).}
\end{figure}

\begin{figure}[h]
\centering
\includegraphics[width=.6\textwidth]{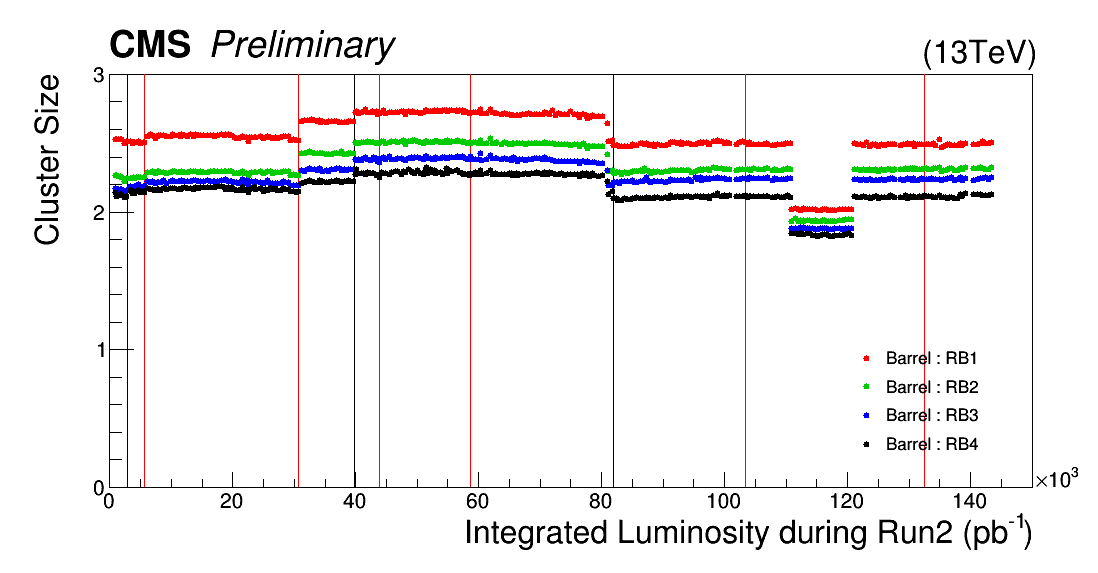}
\caption{\label{BarrelCLS}RPC average cluster size vs integrated luminosity during Run-2 for barrel stations. Red vertical lines show the planned technical stops (TS) and the grey ones — Year-End-Technical stops (YETS).}
\end{figure}

The Figures \ref{BarrelEff} and \ref{BarrelCLS} shows the Run-2 efficiency and CLS history. Each point corresponds to the average efficiency or CLS per barrel station. Each change in the trends is related to the deployment of a new working point. With exception to a drop in efficiency and CLS between in August 2018 caused by a configuration setting issue, it is possible to see that the RPC system had a stable performance with efficiency greater than 95 \% and CLS between 2 and 3, within the CMS Muon Trigger requirements.

\subsection{RPC contribution to the CMS Muon Trigger}

The main contribution of the RPC system is to the CMS Level-1 (L1) Muon Trigger \cite{L1Trigger}, which is divided into 3 $\eta$-regions: the barrel muon track finder (BMTF) for $|\eta| < 0.83$, the overlap muon track finder (OMTF) for $0.83 < |\eta| < 1.24$ and the endcap muon track finder (EMTF) for $|\eta| > 1.24$. Each one of the track finders can access information of the detectors concurrently, building tracks and assigning p$_T$ and exploiting the redundancies of the muon system. 

\begin{figure}[h]
\centering
\includegraphics[width=.4\textwidth]{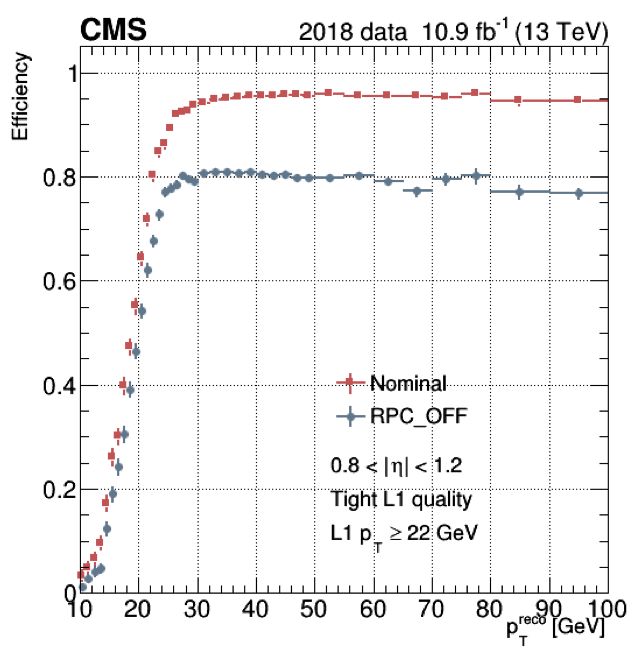}
\caption{\label{OMTFEff}Trigger efficiency versus p$_T$ for the OMTF derived from algorithm emulation applied on real data. Using (Red) and not using (Blue) RPC information. Ref. \cite{RPCL12020}.}
\end{figure}

In the BMTF, RPC timing information is used to improve DT trigger primitives and bunch crossing assignment. In the OMTF, RPC-only segments are build from the 8 chambers available, resulting in a substantial gain in in efficiency. Figure \ref{OMTFEff} demonstrates that using RPC information in the OMTF increases efficiency by 15 \%. Finally, in the EMTF, RPC hit position is used in case of CSC trigger primitive absence. More information on this subject can be found in \cite{RPCL12020}.

\subsection{RPC ohmic current monitoring}

The ohmic current is defined as the current with no beam, up to around 7000 V, where the gas amplification contribution is negligible and the currents follow the Ohm's law.

\begin{figure}[h]
\centering
\includegraphics[width=.8\textwidth]{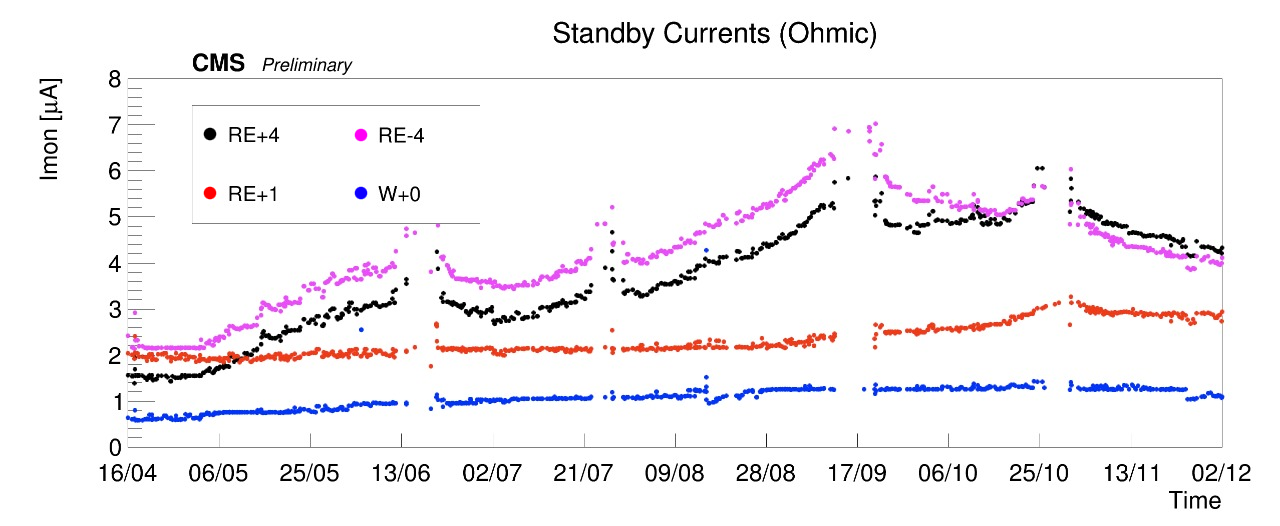}
\caption{\label{ohmiccurrentvstime}RPC ohmic current monitoring history for the stations W+0, RE+1, RE+4 and RE-4.}
\end{figure}

Figure \ref{ohmiccurrentvstime} shows the history of ohmic current measured in four RPC stations. An increase in the currents was observed in all stations, correlated to the background, the RE+4 and RE-4 background rate is about 40 Hz/cm$^2$ and W+0 and RE-1 is less than 10 Hz/cm$^2$. In November of 2018, at the start of the Heavy Ion period where the luminosity is very low and the background is effectively zero, the currents started to decrease. In RE-4 the decrease was faster with the increase of the gas flux.

The increase was correlated with the production of flourine ions (F$^-$) in the gas gap due to the electrical discharge inside the gap. The flourine can combine with the water in the gas and form HF, that can damage the gap. This explains the velocity of decrease of currents correlation with gas flux, as the gas can flush out the F-ions before the combination with water. More information on the fluorine formation inside RPC gas gap can be found in \cite{HF}.

\section{LS2 Activities}

During LS2, CMS is undergoing an intensive upgrade and maintenance program during the second two-year-long shutdown period (LS2). To ensure an excellent performance of the detector in the subsequent physics program, the RPC group pursued in the present shutdown a thorough detector consolidation program to repair most of the hardware problems. In preparation for future installation of new iRPC chambers, cooling and cable services for the new detectors were installed. This includes thousands of kilometres of high voltage and low voltage cables, stainless steel gas pipes between predistribution gas racks in the Service cavern (USC) and gas distribution racks in the experimental cavern (UXC), copper pipes between distribution racks and chambers, gas impedance boxes, support equipment, and optical fibers for the iRPCs.

In order to keep the optimal performance of the system, an extensive HV and low voltage (LV) maintenance campaign was performed. The goal of HV maintenance was to identify the problematic parts of the HV power system and to fix it in the best possible way, recuperating the performance of the chambers. A total of 65 HV channels were repaired. The LV maintenance aim was to ensure a proper operation and configuration of the detector electronics and ensure a good functionality of the LV power boards and communication buses. A total of 12 LV problems were fixed.

A very important activity was the extraction of the chambers from the two RE4 stations to allow the CSC ME4/1 chamber extraction for electronics refurbishment. The chambers were brought to the surface, and accommodated to a new laboratory with controlled environmental conditions and new gas lines to provide the standard RPC gas mixture. All needed reparation and re-validation was done in the laboratory before the re-installation.

The activity with most priority during LS2, was gas system consolidation. The aim was to minimize the environmental impact of the RPC system, as the standard gas mixture is mainly composed of fluorine composed gases (F-gases) with high global warming potential (GWP). The actions taking place during LS2 are:
\begin{itemize}
    \item Gas Leak identification and repair, where 49 out of 99 leaky chambers in barrel where repaired.
    \item Recuperation of the Exhaust, which was not working during Run-2 for the installation of the first $C_2H_2F_4$ recuperation system with efficiency of 80 \%, which have been developed by CERN EP-DT Gas team. 
    \item By end of 2021, CERN EP-DT Group is going to install automatic pressure regulation valves on the redistribution gas racks in USC to minimize pressure variations in the chambers, which can be a possible source of new leaks. 
    \item Turn off the remaining leaky chambers which was not possible to repair (about 3.5 \% of RPC system), to keep the amount of fresh gas added to the system at a minimum.
\end{itemize}

\section{RPC commissioning with cosmic rays data during LS2}

\begin{figure}[h]
\centering
\includegraphics[width=.6\textwidth]{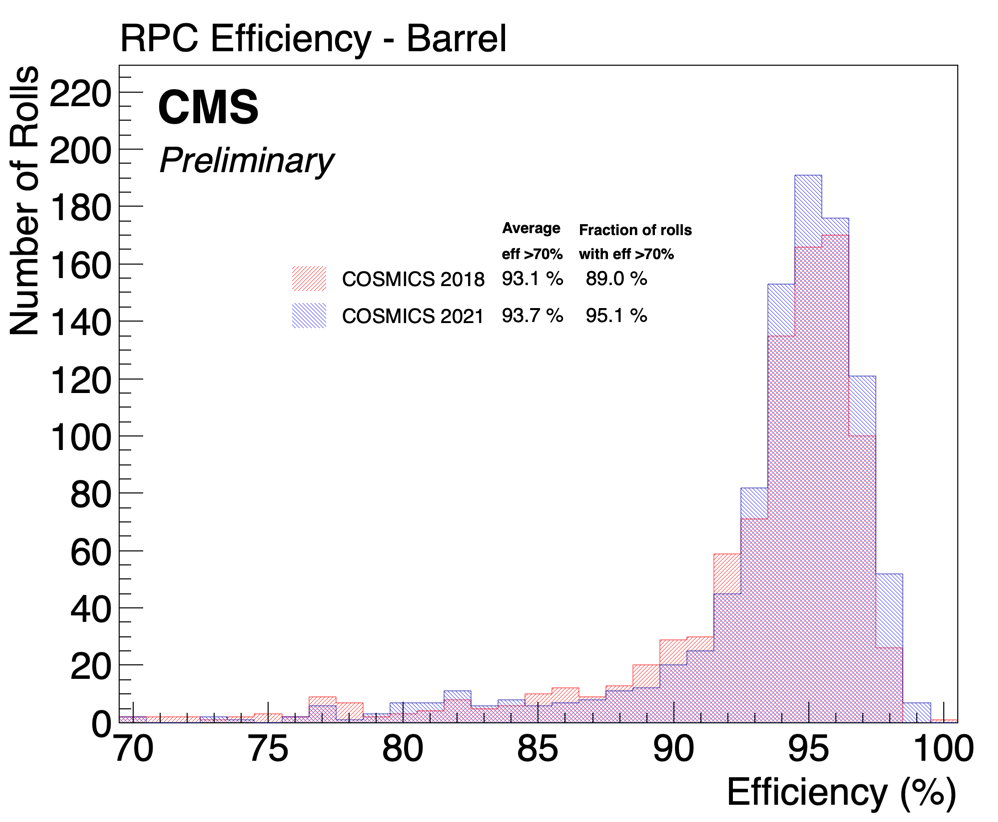}
\caption{\label{effcomp}RPC barrel efficiency comparison between cosmic runs of 2018 (Red) and 2021 (Blue).}
\end{figure}

In Figure \ref{effcomp}, the efficiency of the barrel chambers is compared using cosmic rays runs of 2018 (End of Run-2) and in 2021 (after all the repairs of LS2). The average efficiency of the chambers with efficiency greater than 70 \% is compared and is about the same (gain of 0.6 \%). The fraction of chambers with efficiency greater than 70 \% increased in around 6 \% as a result of all the repairs, which is very important for the muon system redundancy. 

\section{Conclusion}

The CMS RPC system showed a stable performance after operation in Run-1 and Run-2 ($\sim$185 fb$^{-1}$) with average efficiency greater than 95 \% and average CLS between 2 and 3 strips/hit. We observed reversible ohmic current increase in high background regions correlated to the F$^-$ formation inside the gaps. During LS2 RPC did a massive maintenance campaign to repair gas leaks, HV and LV in order to keep the optimal performance and is planning to keep the greenhouse gases emissions at a minimum by keeping all the remaining leaky chambers off and using the new $C_2H_2F_4$ recuperation system. The performance of the RPC System is improved with respect to the end of Run-2 and the system is ready and commissioned for Run-3 data taking.

\end{document}